# Making Mobile Augmented Reality Applications Accessible

Jaylin Herskovitz*
University of Michigan
jayhersk@umich.edu

Jason Wu*
Carnegie Mellon University
jsonwu@cmu.edu

Samuel White
Apple Inc.
samuel_white@apple.com

Amy Pavel
Apple Inc.
apavel@apple.com

Gabriel Reyes
Apple Inc.
gareyes@apple.com

Anhong Guo
Carnegie Mellon University
anhongg@cs.cmu.edu

Jeffrey P. Bigham
Apple Inc.
jbigham@apple.com

## ABSTRACT

Augmented Reality (AR) technology creates new immersive experiences in entertainment, games, education, retail, and social media. AR content is often primarily visual and it is challenging to enable access to it non-visually due to the mix of virtual and real-world content. In this paper, we identify common constituent tasks in AR by analyzing existing mobile AR applications for iOS, and characterize the design space of tasks that require accessible alternatives. For each of the major task categories, we create prototype accessible alternatives that we evaluate in a study with 10 blind participants to explore their perceptions of accessible AR. Our study demonstrates that these prototypes make AR possible to use for blind users and reveals a number of insights to move forward. We believe our work sets forth not only exemplars for developers to create accessible AR applications, but also a roadmap for future research to make AR comprehensively accessible.

*Also with This work was done while Jaylin Herskovitz and Jason Wu were interns at Apple Inc.

## 1 INTRODUCTION

Augmented Reality (AR) has proven useful in a wide variety of application areas, such as retail, education, and social media. Although AR content can be audio-based [8, 25, 39], AR is often primarily visual, and thus making this content accessible non-visually is challenging. Prior work has considered how to make Virtual Reality (VR) accessible [72], which is a related but very different problem. In VR, the entire immersive environment is generated computationally. It is thus conceivable to build in semantics that would allow the entire virtual world to be accessible (e.g., [34]).

In contrast, AR adds virtual content into the real world. Sometimes that content is only overlayed and is thus easily separable from the physical world (e.g., Google Glass notifications). However, oftentimes, AR applications involve actions and objects that bridge between the physical and virtual world, such as scanning the space to initialize the AR model, or placing virtual objects in relation to both other virtual objects and real-world objects. Such applications are especially difficult to make accessible since doing so requires knowledge of not only the virtual objects the application creates but also the physical context into which they are placed.

AR technologies have also been explored in the context of improving accessibility to the real world. For instance, CueSee uses head-mounted AR to help people with low vision better find objects of interest by visually overlaying different cues to help mark an item or make it easier to see [77]. VizLens overlays an audio interface onto visual (and inaccessible) physical interfaces [22], so that a blind person can use them. In this paper, we consider the different problem of how we might enable developers to make existing AR applications, which are not specifically designed in advance for non-visual interactions, possible to be used non-visually.

As such, our work builds on a long history of accessibility work, which has introduced technological means to make visual computer interfaces accessible in other ways. For instance, screen readers have been developed to make graphical user interfaces and windowing systems accessible [6, 44]. We believe we are now at a critical time in the development of AR, where we can think ahead about how to make sure that AR applications are accessible to everyone as they are emerging [2, 42].

Applications are being developed using AR for a wide variety of innovative and interesting reasons. An alternative approach could be to not use AR for content that needs to be accessible, or to create separate versions of the content that does not use AR. However, as has been shown again and again in the history of accessible technologies, separate more accessible versions of software rarely provide an equal experience. Instead, content is slower to arrive and becomes out of date, and functionality is limited and not maintained to the same degree as the application it is intended to parallel [68]. In this paper, we explore how we might make existing AR applications natively accessible so that everyone can benefit from them.

Understanding the design space of tasks in AR is an important first step to designing accessible alternatives. Accordingly, we first collected and thematically grouped the interactions required to use 105 existing mobile AR applications. We chose to focus on smartphone AR applications in this work because the smartphone platform is nearly ubiquitous, and smartphone AR is quickly finding its way into a number of important applications. From this analysis, we present a design space for existing constituent tasks found in AR applications, which we believe can serve as a roadmap for future research and development in making these applications accessible. While we focus on AR applications for mobile phones, we believe our results can inform the design of AR applications for a variety of smartphone and head-mounted platforms, which all are facing the same challenge of how to make themselves accessible. We identified five key categories of tasks within AR applications and described them along with examples and considerations that affect accessibility for each.

Using this design space, we then selected three of the most common tasks and designed prototypes of accessible alternatives for each to serve as design probes: *(i)* scanning the environment to initialize the AR world model, *(ii)* placing virtual objects in the space, and *(iii)* locating and exploring virtual objects in the space. We also created two full AR apps combining these components that were meant to mimic common AR use cases: a retail app designed to allow the user to explore how furniture might fit into the context of their own environment, and an educational app in which users could explore the solar system. We then conducted a user study with 10 blind participants to gather feedback about each design.

The main contributions of this work are:

(1) A taxonomy of constituent tasks found in 105 existing AR applications, which provides a roadmap for future research in making AR applications accessible.
(2) Five exemplar prototypes of non-visual alternatives to common AR tasks, and two prototypes that combine them into realistic full AR applications that are non-visually accessible.
(3) A study in which we used our design probes to explore how 10 blind participants interacted with AR on mobile devices.

## 2 BACKGROUND

Mixed-reality systems exist on a continuum between the real and virtual world [40, 41]. Our work focuses on augmented reality (AR), which introduces virtual elements into the real world. Our work builds from work on *(i)* 3D and mobile applications accessibility, *(ii)* camera-based applications for making the world more accessible, and *(iii)* defining the capabilities of mixed-reality systems.

### 2.1 Making Applications Accessible

Mobile applications, like their desktop and web analogues, can be made accessible by following application guidelines, such as the UIKit Accessibility Guidelines [4]. These best practices are related to guidelines such as the Web Content Accessibility Guidelines (WCAG) [66], which provide guidance on how to provide the semantic information necessary to make the content of a user interface accessible. Many mobile applications across different platforms are insufficiently annotated to be fully accessible [54], which has led to various attempts to improve their accessibility after the fact, e.g., through run-time repair [70]. Additionally, as smartphone hardware has changed, the assistive technologies that operate on devices have adapted, for example, with new approaches for enabling a person who is blind to use a touchscreen interface [30].

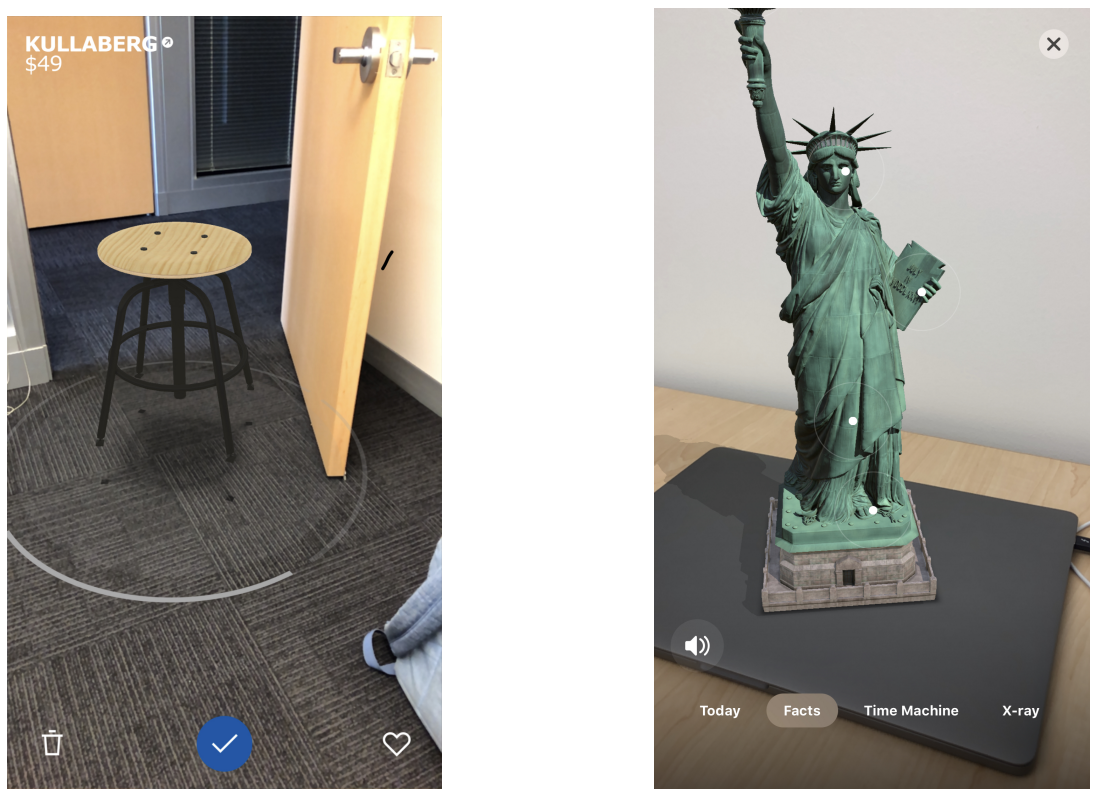

**Figure 1: Left: *IKEA Place* [28] allows users to view furniture in AR. Right: *Statue of Liberty AR* [60] displays historical facts along with to-scale models of the Statue of Liberty.**

Relatively little work has considered how to make the AR applications that are quickly becoming popular more broadly accessible, aside from guidelines on the use of color and audio in AR for users with low vision or hearing impairments [37]. Prior work has considered how to make VR accessible for people with visual impairments via audio [67] and haptic feedback [32, 72]. For example, SeeingVR introduces methods for making VR accessible to low-vision users, and largely takes inspiration from prior systems for providing access to the digital and physical worlds, e.g., through adjustments to visual content and through various automated methods for describing or enhancing the virtual content at runtime [18, 72, 76]. Other work looks to leverage the advantages of technologies, like the white cane, with which some people with disabilities are already familiar in order navigate virtual content [58, 64, 71]. Prior work has also considered how to make other 3D applications, such as games [3, 19] or CAD software [56, 57], accessible through similar techniques. In our work, we instead consider AR, which differs in that it is a combination of real and virtual content. We also identify common constituent tasks in AR applications and provide patterns for how those might be made accessible, which we hope will be useful for developers.

### 2.2 AR for Making the World More Accessible

AR has also been used to improve accessibility across a wide variety of domains, such as for visual assistance for people with low vision [21, 73, 75] and color blindness [62], assistance for people with cognitive impairments [31], and coaching for rehabilitation [11]. As examples, CueSee [77] and ForeSee [76] enable visual identification and search for items of interest, respectively. AR has also been used to help people with low vision better navigate on stairs [74]. In general, these systems seek to augment users' perception and cognitive abilities via visual overlays in their real environment. Few of these systems have used the smartphone platform, opting instead for a head-worn form factor or augmenting the user's environment.

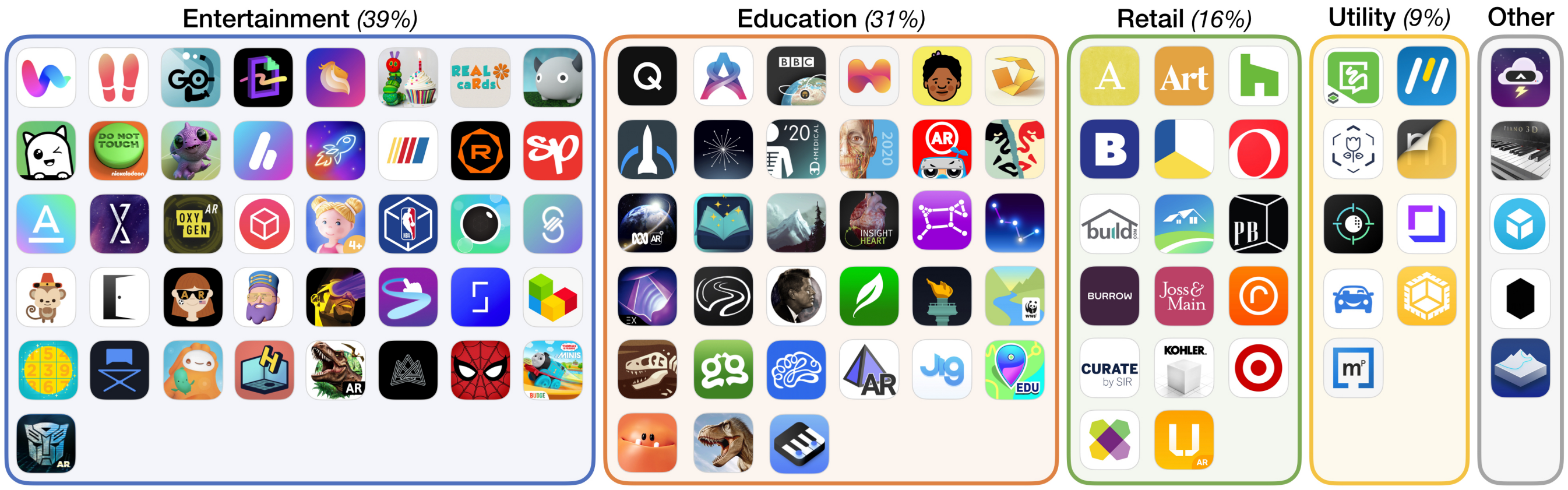

**Figure 2: Icons for the 105 AR apps that we analyzed, organized by category.**

Technologies associated with AR have recently been used in a variety of systems intended to improve access to the world using smartphones. A number of systems have been developed to help blind people take better photos, generally by using automated approaches to assist in aiming the camera [29, 36, 65]. VizLens uses a combination of computer vision and crowdsourcing to recognize and guide a blind user through using an inaccessible physical interface [22, 23]. Cursor-based interactions assist blind people to attend to and interact with physical objects in complex visual scenes [24]. Audio-based AR systems such as Microsoft's Soundscape [39], Blindsquare [7], and NavCog [1, 55] act as navigation aids which provide information on points of interest and non-visual landmarks. These systems do not introduce visual augmentations but rather augment the environment with audio cues that help users perform tasks of interest in the real world. This work helped to inform the design of our accessible AR prototypes.

## 2.3 The Space of Mixed-Reality Systems

Existing taxonomies of AR/VR systems [40, 59] mainly describe systems in terms of what capabilities the display hardware affords. Other analyses of immersive software are largely focused on specific domain areas, such as industrial manufacturing [12], games [14], automotive applications [69], or medicine [20]; specific functionality, such as visualizing relationships between physical and virtual information [43]; or specific interactions, such as gestures, gaze [26], or using a tablet alongside a VR headset [17, 61]. The results of an extensive AR gesture elicitation study demonstrate that the space for gestures in AR is quite large [52]. This prior work primarily focused on the form of gestures and how participants thought they should map onto a pre-defined set of tasks (e.g., select, open, delete-x). We instead consider what "tasks" need to be done to fully make use of common AR applications.

# 3 DESIGN SPACE OF AR INTERACTIONS

A precursor to creating accessible alternatives to current AR interactions is understanding what tasks are currently common and necessary in AR applications. Such an understanding is also a prerequisite for usability evaluation [47] and modeling [13]. This is also implicitly related to how assistive technologies have been developed to work on graphical user interfaces (GUIs) – tasks are identified first and then accessible alternatives to them have been invented. For instance, "drag and drop" was introduced, and then an accessible way to perform the same function was developed and introduced. While common tasks are largely known and repeated in GUIs, interactions in AR are much less explored, and AR affords even greater flexibility on what interactions are possible.

Our goal was to discover repeated constituent tasks across different applications so that we could then develop approaches for making them accessible, thus providing developers useful patterns that they could follow to make their own applications accessible. We performed an analysis of the functionality and design of existing mobile AR apps, and from this, we present a description of the design space of AR apps and a set of common constituent tasks.

## 3.1 Dataset

Our dataset consists of all AR apps that were displayed on the 'AR' category page of Apple's App Store for iPhone over a three month period (June to September 2019). Two examples are shown in Figure 1. We removed apps that did not have apparent AR content, as well as one with location-specific content we could not access, leaving us with 105 apps. Of these apps, 83 (79%) used AR as the main feature of the app, while the remaining 22 (21%) used AR as a secondary or supporting feature. The apps that we evaluated were spread over a variety of categories in the App Store, which we further condensed into the following five groups: 39% Entertainment, 31% Education, 16% Retail, 9% Utility, and 5% Other. An overview is shown in Figure 2. Our analysis focused on the iOS platform, as many people with disabilities are iPhone users, including those who are blind and those who use switch control. AR support on other popular platforms, such as Android, is similar, although hardware is more varied. Given that many of the apps we analyzed are available on both platforms, we expect our findings to generalize.

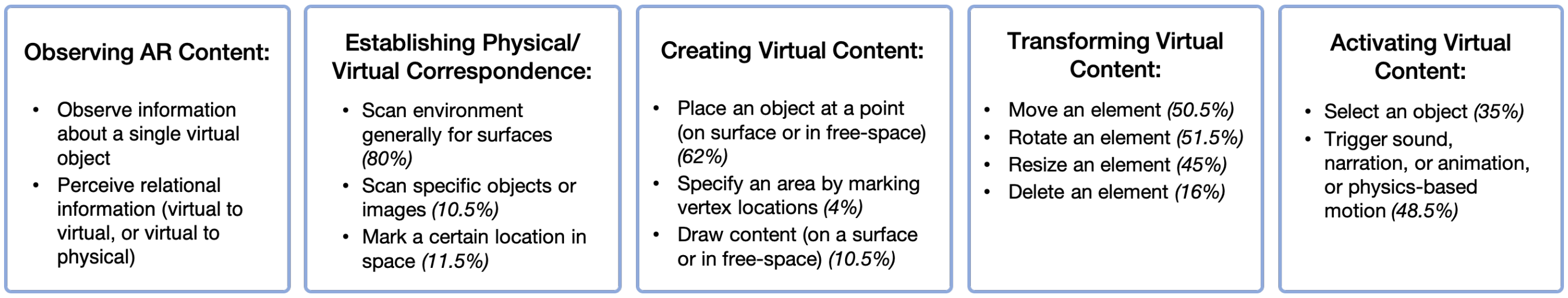

**Figure 3: A summary of the identified 'building-block' tasks needed to interact with AR apps, and the percentage of apps that contained each task, with higher-level categories identified.**

## 3.2 Methods

Three members of the research team analyzed the apps in our dataset using thematic analysis [10]; all had experiences with AR/VR systems, accessibility, and/or qualitative coding. We followed the six phases that Braun and Clarke described [10], treating screenshots and textual descriptions of app functionality as data items. Each researcher performed the first two phases individually, analyzing a subset of 15 apps and performing an open coding of their observations. The remaining phases were completed as a group; the research team iteratively adjusted the codes using this process until sufficient agreement was reached. We used Randolph's free-marginal multirater kappa [53] to measure agreement. We considered a code finalized when we reached a kappa value of 0.7 or higher. An overview of the resulting codes is shown in Figure 3, with frequency of occurrence by app category shown in Figure 4.

## 3.3 'Building-Block' Tasks in AR

*3.3.1 Observing AR Content.* The type of visual AR that exists on modern smartphones gives the appearance of three dimensional, virtual objects that have been placed in the physical environment. Thus, in order to perceive all aspects of the virtual content, users are required to 'look around' the space, using the phone's camera as a lens by which to view the virtual world. This serves multiple purposes, just as visual perception does, including perceiving information about a single virtual object (its size, shape, color, style, etc.), as well as information about the relationships between objects. Relational information can be between a virtual object and its physical surroundings (e.g., to check if a virtual product matches one's home) or between multiple virtual objects (e.g., to compare the size of two virtual products), and includes how virtual objects may be similar or different in appearance and how they are arranged in the physical space. This task also serves to enable users to discover new content and functionality. For example, in *Forensic Detective* [46] the user must search for hidden clues around their room and move the phone close to virtual content in order to interact with it.

All of the information above needs to be conveyed in an alternative form for visually impaired users. As blind users typically familiarize themselves with physical spaces through their sense of touch, this task is difficult to replicate at the same level of fidelity without additional haptic devices. For example, the Canetroller [71] is a device that simulates white cane interactions and provides physical resistance and vibrotactile feedback for objects in VR. Designing accessible interactions which can convey this information using commodity smartphone hardware presents a challenge.

*3.3.2 Establishing Physical/Virtual Correspondence.* A necessary class of tasks within AR involves creating a relationship between the physical space and virtual content in order to create a basis for positioning virtual content. Current smartphone AR systems usually use an RGB camera to identify visual features and detect physical surfaces in the space, and require users to pan the camera slowly to do so successfully. 15.2% of apps in our dataset did not require any of these tasks, and instead placed content relative to the position of the camera only.

80% of the apps in our dataset asked users to perform a general scan of their space to detect surfaces. In order to perform this scan successfully, users need to first be aware of how the software expects the phone to move, which is usually explained through animations. Users also need to know if the environment has sufficient lighting and sufficient visual features for detection, something that modern systems can notify the user of. There are also application dependent factors that users need to be made aware of, for example, how large of a surface the app requires, or if a specific type of surface is required (i.e., table or floor).

Establishing virtual/physical relationships can also be more explicit. 10.5% of the apps in our dataset asked users to scan a specific object or image to annotate with virtual content. For example, *Tonic* [15] asks the user to scan a piano to show chord information, and *Waypoint EDU* [38] will recognize pre-printed posters placed in a classroom as triggers to display educational content.

Additionally, in some cases where the app is unable to recognize a point through other means, the user is asked to specifically label a point or an area by placing a virtual marker in the space (11.5% of apps). For example, in *TapMeasure* [48] the user is asked to mark the corners of their room with small flags in order to measure the length of each wall, and in *Tonic* [15], if a piano is not automatically detected, the user is asked to mark the first and last keys with virtual dots in order to align content. This interaction represents a failure mode of AR tracking that would be much more difficult for a visually impaired user to recover from, as existing implementations require precise camera alignment to mark specific points.

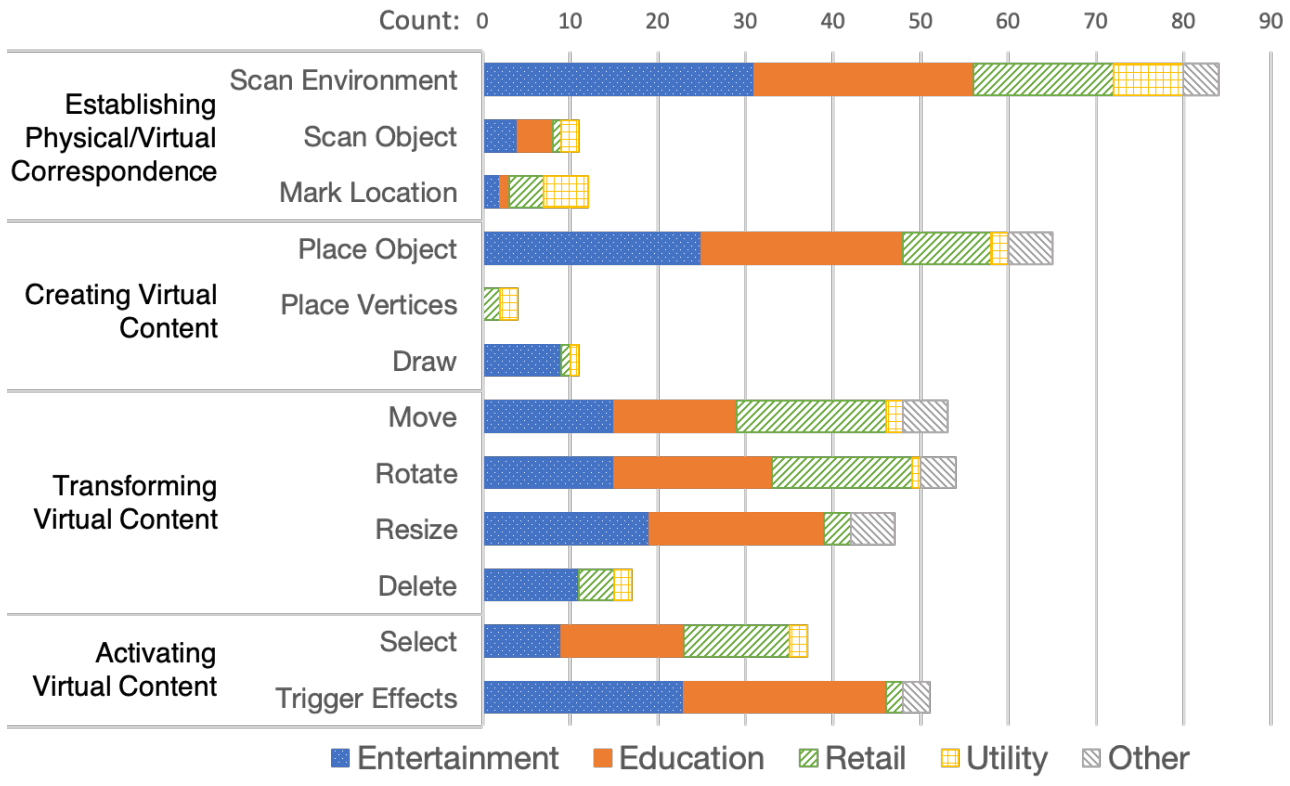

**Figure 4: A breakdown of each constituent task we identified and how frequently it appeared in each app category.**

*3.3.3 Creating Virtual Content.* 70.5% of the apps in our dataset allowed users to place virtual content in some way, while the rest placed content automatically for the user. We identified three common modes of placement: (1) by indicating a specific point at which to center the object (62% of apps), (2) by indicating a series of points to form a polygon to fill with content (4% of apps), or (3) by drawing free-form lines by dragging their finger on the screen (10.5% of apps). In order to place virtual content, users need to evaluate both *possible* locations (i.e., locations that would be able to accommodate the size of the item), and *appropriate* locations (i.e., locations that make reasonable sense for the item, for example, a virtual chair should not be placed on a table). An app should provide sufficient information for users to make this evaluation.

*3.3.4 Transforming Virtual Content.* Traditional 3D manipulations are also extremely common in AR apps, with 68% of apps in our dataset allowing at least one of the following forms: (1) editing position (50.5%), (2) editing orientation (51.5%), (3) editing scale (44.8%), or (4) deletion (16.2%). In cases where the position or orientation of an object could be edited, this was usually constrained to two dimensional motion along a surface or rotation around one axis, as described in Apple's design guidelines for AR apps [5], as more complex manipulations are difficult with touch controls.

The goal of each manipulation is highly dependent on the context: one may rotate a piece of furniture so that it fits in their room (crucial), rotate an educational model to see what it looks like from another angle (optional, but they may learn additional information), or rotate a 3D emoji because they like the way it looks (purely cosmetic). Additionally, while performing a manipulation, users often need to observe other parts of the scene at the same time. For example, when resizing an object, sighted users can compare its size to other virtual and physical objects in the area, as well as to the size of the physical space in general, in order to determine what is appropriate. Alternative mechanisms for users with visual impairments must also convey this information.

*3.3.5 Activating Virtual Content.* Users are often required to select a specific object in the scene, which may trigger additional effects. This can be simple, such as selecting an object so that it becomes editable or additional text information about the object is displayed (35.2% of apps), or can include more complex effects such as sound effects, animations, or physics-based motion of an object (48.6% of apps). This category is by far the most diverse. Even the distinction of complex effects can range from localized animations which may just need audio descriptions, to game mechanisms, such as swiping a finger across the screen to toss a virtual basketball into a net as in *NBA AR Basketball* [45]. In general, an app should provide users with sufficient information to determine what objects can be activated, their current state, and the resulting effect on the scene.

# 4 PROTOTYPES OF ACCESSIBLE AR

The prior section introduced the taxonomy of constituent tasks that we identified from existing AR applications. In this section, we explore how such applications could be made accessible. We believe that creating truly accessible alternatives to many different constituent tasks and the experiences that they embody is a long-term research task. Instead of attempting to solve the whole problem, our goal was instead to *(i)* demonstrate that common AR tasks and applications can be made accessible, and *(ii)* create prototype accessible AR applications for use in the studies with blind participants that conclude this paper.

We first present foundational work for exposing virtual objects displayed in AR to the users of accessibility services, such as screen reader and switch control users. While a number of AR applications exist across mobile platforms, we developed our prototypes for iOS and specifically targeted VoiceOver use.

We then present five constituent task prototypes that we developed to illustrate how AR applications might be made accessible: one for scanning surfaces (from the set *Establishing Physical/Virtual Correspondence*), two for placing virtual objects on surfaces (from the set *Creating Virtual Content*), and two for locating virtual objects in the space (from the set *Observing AR Content*). In each of the two latter cases, we drew from prior work to create contrasting experiences: one experience attempted to directly make the existing experience accessible, and the other experience attempted to assist the user in performing the task in an alternative way.

The tasks we prototyped were chosen for their ubiquity, as well as their ability to combine to form realistic, full AR applications. We present two such full apps, in the domains of retail and education. These are common applications for smartphone AR, at 36% and 19% of our dataset respectively. Retail was chosen over the slightly more common entertainment category because of the higher level of precision needed to place and evaluate virtual products.

## 4.1 A Foundation for Accessible AR

The first step in making AR content accessible is to make accessibility services aware of virtual content and allow developers to assign metadata to it. To this end, we added a bridge that exposes the underlying structure of the AR scene to VoiceOver, making each side of a 3D object's bounding box the same as any other 2D UI element on the screen, similar to the touch cursor mode in prior work [24]. Our approach was implemented on the SceneKit objects that ARKit uses to add virtual objects by adding them to the view hierarchy that accessibility services use to traverse applications. As a result, our approach is general and can apply to many existing applications that use this toolkit for creating AR experiences. Transforming 3D content into 2D targets provides a foundational awareness of what content is in an AR scene and functionality for selecting objects akin to 2D buttons (Figure 5A).

We also implemented a "freeze" feature that captures a stable view of the AR content and the physical world it is overlayed onto, enabling users to interact with the frozen view without worrying about moving the device. We found this to be an important usability feature because otherwise blind users of AR needed to keep their mobile devices positioned such that they always point at the virtual objects of interest. Aiming cameras non-visually is, in general, known to be a hard problem [29], and we found that the difficulty is only magnified when the position must be held stable while also interacting with the mobile device. Our "freeze" feature is toggled using the two-finger double-tap ("magic tap") in VoiceOver.

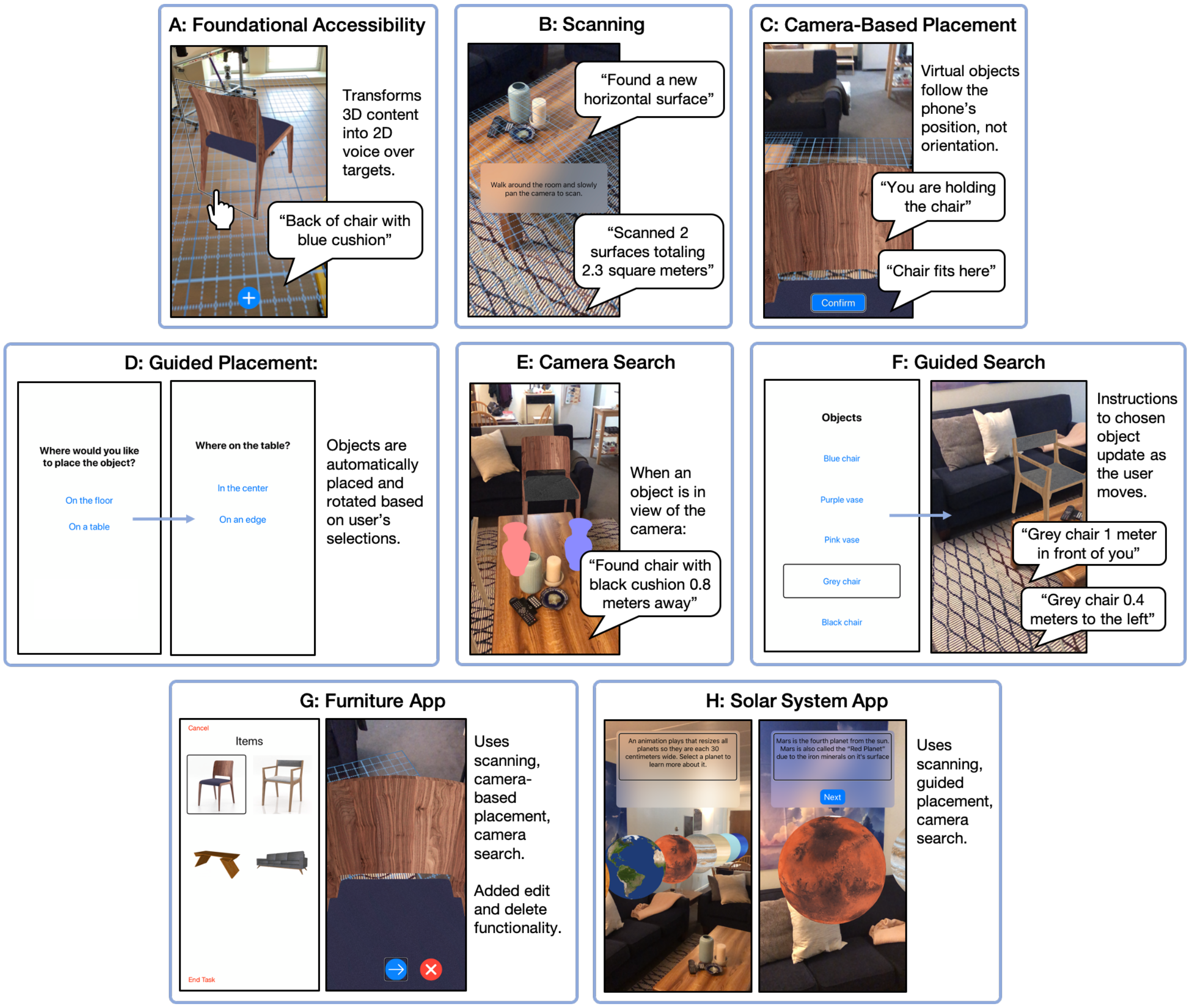

**Figure 5: Our approaches to making AR apps accessible: the foundational level of accessibility added to VoiceOver (A), five prototypes for three constituent tasks (B-F), and two apps which combine the tasks to create full AR experiences (G-H).**

## 4.2 Scanning the Physical Space

Our scanning prototype adds a set of accessibility notifications to increase awareness of the current scan progress (Figure 5B). When the user scans a new surface, the app announces this and gives the type (horizontal or vertical). As the user scans more area, the app will periodically announce the number of surfaces and total area that has been detected. Finally, if the user has not scanned any new area in the past five seconds, the app will instruct the user to move to a new area to scan.

## 4.3 Placing Virtual Objects

*4.3.1 Version A: Camera-based Placement.* Camera-based placement uses the position of the phone to determine the position of virtual objects, regardless of the device's orientation (Figure 5C). As users walk around the room, the object is placed on the floor at their feet, and moves as they move. If they hold the phone above a table or other surface, the object will move to that surface and continue to follow the phone's position; the user will receive a verbal notification when this occurs. As the user walks around their space with the object, they also receive notifications when the object does not fit in the area they are standing, for example, if it is too close to a wall, too close to another virtual object, or too large to fit on or underneath a table. When the user is ready to finalize placement, they press a 'confirm' button to drop the object in its current location. As the location of the virtual object is tied closely to the user's physical position, this could help give users a sense of where virtual objects are located. However, it also requires somewhat precise movement from the user, as well as some degree of awareness of where the phone is generally pointing, as this is used to orient the object so that it appears to face the user (though it is not used to position the object).

*4.3.2 Version B: Guided Placement.* Guided placement generates a series of candidate positions for an object based on what surfaces have been detected, then asks the user a series of questions about where they would like the object to be placed in order to determine the best position and orientation (Figure 5D). Users are first asked if they would like to place an object on the floor or on a table. Based on their selection, they are then provided with another series of options: if table was selected, they are asked to choose between the center of the table or an edge of the table; if the floor is selected, they can select between the center of the floor, an edge, or a corner of the room. If applicable, users are then asked to face the edge, wall, or corner that they would like the object to be placed against. For example, if the user would like to place a chair against a wall, the object can then be positioned so that it is exactly against the wall and is rotated correctly. This constrains placement as compared to the location-based method, as only a subset of positions in the room are detected as candidates for an object. However, this option requires less work for the user in determining where an object would fit (as this is determined by the app) and how to best rotate it in the space. Thus, it is likely better suited for cases where the exact placement of a virtual object may not be as important to the user. Additionally, content was placed on the first table scanned as current mobile AR can detect only horizontal and vertical planes; future scanning mechanisms could enable other targets to be presented as options.

## 4.4 Finding Virtual Objects

*4.4.1 Version A: Camera-based Search.* Using camera-based search, users scan the camera around their space to find objects, similar to the window cursor mode in prior work [24]. When the user points the camera at a virtual object, they receive a verbal notification stating the name of the object and how far away it is (e.g., "Found chair 0.5 meters away"), and will also feel a haptic vibration from the phone (Figure 5E). When they move the camera away from the object they receive a similar notification. Using this information, the user can walk in the direction that the camera is pointing to locate a found object. When they are close to an object (within a certain threshold), there is an additional notification; in this way the user can get a sense of the locations of virtual objects. Although this requires the user to point the camera at the location of a virtual object, it could allow for a more free-form exploration of the space, which may be useful in some applications.

*4.4.2 Version B: Guided Search.* Guided search presents users with a list of all objects in the space around them, sorted by how close they are to the user's current position. When the user selects an object from the list, the phone then issues verbal directions which update every three seconds as the user moves (Figure 5F). For example, the user might hear the following series of instructions: "The chair is 1 meter in front of you", "The chair is 0.5 meters in front of you", and "The chair is 0.2 meters to the left". The directions "forward", "backward", "left", and "right" are approximations chosen based on which is nearest to the object, so the path to the object is not always the most direct path possible, but the position is eventually reached. Users receive a notification when they are close to an object, as before. This option could be easier to use than the camera-based search, but in some situations could come at the cost of freely exploring the room. For example, when using camera-based search, the user could perform a sweeping scan to quickly get a sense of what objects are around them, which would not be possible with this method.

This method is also similar to the guidance mode in VizLens, which gives users directions to reach buttons on an inaccessible physical interface, while our camera-based search is similar to the feedback mode in VizLens, which announces which button a user is near [22]. The evaluation of VizLens found that users preferred guidance when they were not yet familiar with an interface's layout, and preferred the feedback mode when they were. Our two search methods could be applied in a similar manner.

## 4.5 Furniture App

We created an AR furniture shopping app meant to mimic existing retail apps that make use of AR to let users see products within the context of their space, such as *IKEA Place* [28], *Overstock* [49], *Houzz* [27], and *Target* [63]. Furniture placement is a compelling application for smartphone AR, requiring constant comparison between virtual and physical content in order to determine if an item fits within a room in terms of size and style. This requires accurate positioning and realistic content.

In our app, users can select from a list of items, and then use the camera-based placement method for placing the object in the room (Figure 5G). Similarly, they receive notifications when an object does not fit in its intended location (i.e., under a table, too close to a wall, or conflicting with another virtual object). The camera-based search method is used to provide users awareness of where existing objects are. When the user stands close enough to an object, they are given options to edit its position or delete the object. While our prototype covers most of the basic functionality included in this type of app, more complex functionalities, such as understanding the spatial relationships between objects, are not explicitly included and depend on the user's mental model of the space.

## 4.6 Solar System App

We also created an educational app meant to mimic existing AR apps aimed at elementary school students, such as *ARcheology* [35] or *Plantale* [16]. Such apps are also compelling; they are usually more engaging or interesting for students to interact with, while also sometimes providing secondary information through AR, such as real-world scale or layout of content. However, these apps also usually include content that the user may not already have a frame of reference for (unlike furniture), and it is less realistic in that it may be stylized and/or at a different scale than expected.

Our app presents some basic information about the solar system (Figure 5H). First, the user is instructed to face an open area of the room to place a model of the planets. Guided placement is used to place the model in front of the user, though options are not provided directly to the user as in this case there is only one object that needs to be placed. The user then navigates through two panels of information about the solar system, and an animation is played which resizes each planet so that they are equally sized, which is described to the user. The user can then select planets using camera-based search to learn more about them.

# 5 USER STUDY

The goal of the user study was to better understand how our prototypes performed both as standalone methods of interaction, and when integrated into full AR apps. Overall, we sought to understand what the strengths and limitations for each prototype were, which methods were better suited to different contexts, and how users perceived virtual content.

## 5.1 Participants and Apparatus

We recruited 10 participants (3 male, 7 female). Among them, 8 were blind and 2 had low vision; 1 was in the age range of 30-39, 2 were in the age range of 50-59, 4 were in the age range of 60-69, and 3 were in the age range of 70-79. All participants were users who rely on screen readers in order to access their devices, and all had experience using a smartphone for at least 3 years (and at most 10 years). All of the participants were iPhone users. All participants reported little to no prior knowledge of AR or VR.

We implemented the seven prototypes as described in the previous section and installed each as a separate app on an iPhone 8 Plus. The study took place in a well-lit office room, approximately 10 feet by 10 feet in size, that contained a table, wall shelves, and a whiteboard. The center of the room was open.

## 5.2 Procedure

Participants were first asked a series of demographic and background questions. Participants were provided the opportunity to familiarize themselves with the study room, and also to adjust the VoiceOver settings on the provided device. Next, participants were given a short description of typical smartphone AR usage, and given an opportunity to ask any questions on this topic.

Participants were then asked to use our prototypes to complete a series of five tasks, as described below. After using each app, participants were asked to rate their agreement with a series of statements on a scale of one to seven (from strongly disagree to strongly agree): "This task was mentally demanding", "This task was physically demanding", "I feel it is easy to use this app", "I feel very confident using this app", and "I had a sense of what virtual objects were in the environment around me and where they were located". Responses are summarized in Figure 6. After rating each statement, participants were asked to describe any challenges they encountered and the most helpful feature of the app, and were given a chance to give open-ended feedback.

Each session took between 1.5 and 2 hours, and participants were compensated with $50 each. The sessions were video recorded, and timestamps of app launch, termination, and certain actions were recorded and used for further analysis.

## 5.3 Tasks

We designed the following tasks based on our analysis of existing AR apps. All of the tasks were completed in the same order. For tasks 2 and 3, which participants completed twice, the order in which they used each prototype was randomized such that half of the participants always used the 'A' version first and the other half always used the 'B' version first.

**Task 1:** Participants were asked to scan at least four surfaces (with at least one vertical surface) totaling five square meters in the study room. The app notified the user when this was met.

**Task 2:** Participants were asked to respond to a series of five prompts by placing a virtual object in the room. Each prompt described a location (e.g., "Place the chair in front of the desk").

**Task 3:** Participants were asked to locate five objects that were randomly placed in the room (two on the table, three on the floor).

**Task 4:** Participants were asked to use our furniture app to choose a few pieces of furniture that fit in the room, and arrange them. The choices were two chairs, a couch, and a coffee table.

**Task 5:** Participants were asked to use our solar system app to learn some basic scientific facts. They were instructed to follow a relatively guided-narration and explore content as they saw fit.

# 6 RESULTS

## 6.1 Task 1: Scanning

On average, participants took 39.1 seconds (*SD* = 21.3 seconds) to complete the scanning task. All participants agreed that this task was easy to perform, though our study was run in nearly ideal-conditions (well-lit area, surfaces with many feature points) so tracking was not difficult to establish. Although figuring out how to hold and move the device was an issue for some participants (P4, P5, and P9 required additional guidance from the study administrator on how to point the camera), others were able to adjust as they received feedback from the app:

> *"Once I was doing it and getting the feedback, then I thought, 'Well, I'm doing it right.' Because it said you got one surface, so I just kept going."* (P2)

Though participants generally thought that the verbal updates were helpful in providing awareness of the current scan progress, they agreed that more semantic information about what physical objects are being scanned would be useful for additional guidance.

> *"I didn't care about the horizontal and vertical planes, because I didn't know how relevant that was."* (P5)

> *"I noticed when I pointed at the table it said 'horizontal' and 'vertical' when I pointed at the wall, but it didn't really tell me how far the table was from me, or how far the wall was from me. If I couldn't see at all I might be nervous about how far I could move."* (P1)

## 6.2 Task 2: Placing Objects

### 6.2.1 *Version A: Camera-based Placement.*

On average, this task took participants 4.3 minutes (*SD* = 2.6 minutes) to complete, and each object took on average 27.4 seconds (*SD* = 23.5 seconds) to place. Participants commented positively on the clear connection between a virtual object's position and their own physical location:

> *"All I had to do was to move [to the location] and place it. I knew when something wouldn't fit, and I backed off and placed it."* (P6)

Even so, finding the intended physical location was challenging for some, and additional guidance about where physical objects are located would be helpful. For example, P7 commented that this system required them to keep the layout of the room in mind more than they normally would, and P8 said *"Placing the object*

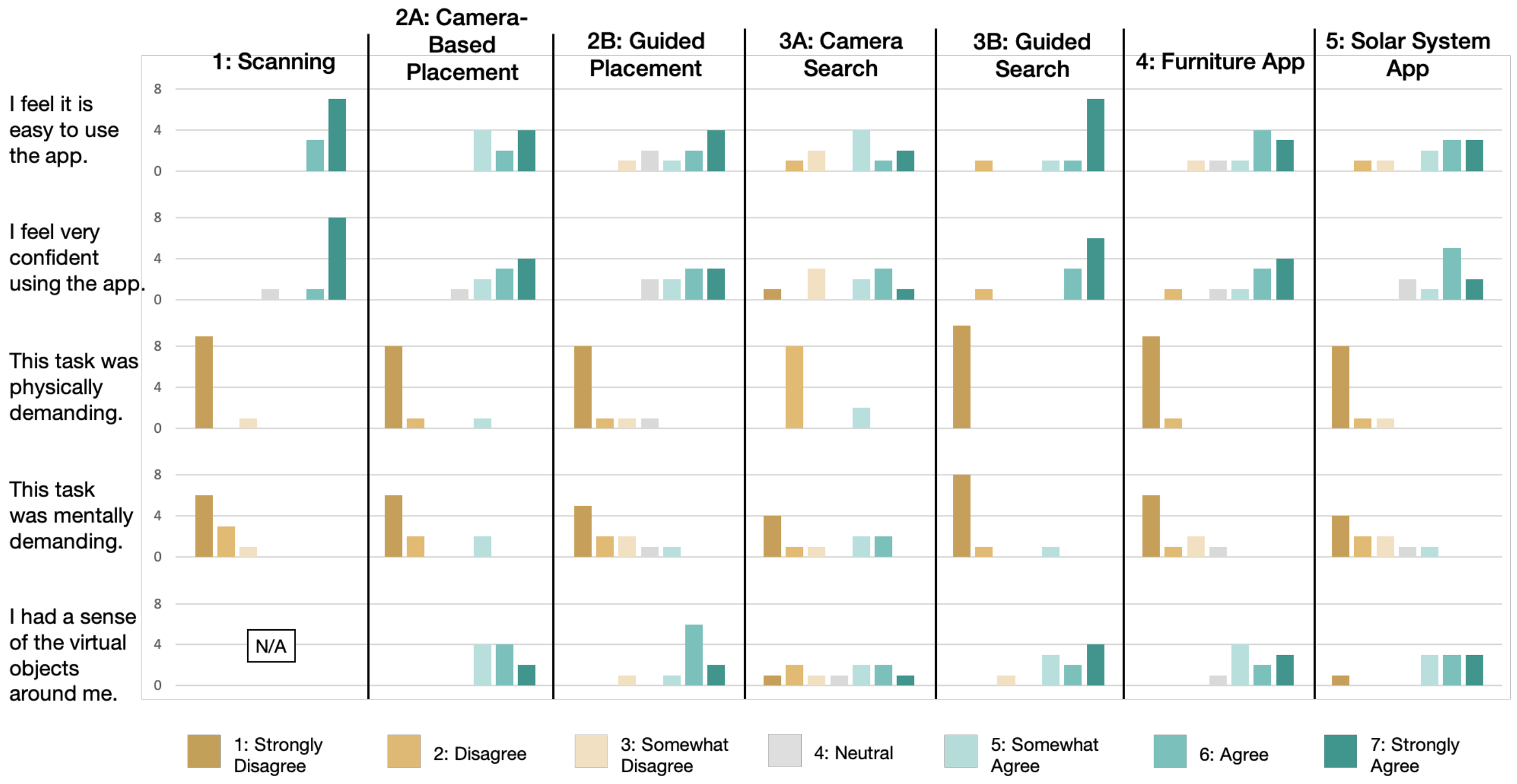

**Figure 6: Responses from Likert scale questions for all tasks. Participants were asked to rate the extent to which they agreed with the statements on the left. Each bar shows the number of participants who gave a specific response.**

*is easy, once I know where I was at."* Additionally, all participants commented that the notifications that an object did or did not fit in their current location were helpful:

> *"It would give me more feedback than [guided placement], like if it didn't fit, you could just move to another corner. As long as it was saying that, I could just find another place. That feedback was really good."* (P2)

*6.2.2 Version B: Guided Placement.* On average, this task took participants 6 minutes (*SD* = 1.5 minutes) to complete. Each object took on average 48.4 seconds (*SD* = 10.5 seconds) to place. Participants noted that the options were easy to navigate, but lacked precision:

> *"I think that one where I had to put the chair in front of the table, but there were only three options, there wasn't enough precision to do what I wanted to do. It wasn't difficult to understand, but there was an imprecision in the placement options."* (P1)

Participants also noted that there was a lack of semantic and contextual information in the options.

> *"It was step by step, that helped. I don't know if it's always necessary to break down all of those little steps... If it would ask you where would you place the chair, it would always be on the floor, so it seemed like some of those steps were needless."* (P3)

## 6.3 Task 3: Finding Objects

Participants preferred the guided search method for finding unknown objects. This was evidenced in the Likert scale responses as well as participant comments, and is consistent with previous findings on user's guidance preferences when presented with an unfamiliar layout [22].

*6.3.1 Version A: Camera-based Search.* On average, this task took participants 5.2 minutes (*SD* = 1.3 minutes) to complete. Each object took on average 68 seconds (*SD* = 15.6 seconds) to find. Of all the tasks, this was the most challenging, as evidenced by participants' questionnaire responses (Figure 6). This is caused by needing to scan the room with the camera when the targets were unknown, which was mentioned by all participants.

> *"If I knew what I was looking for it would have been a lot easier. For the purple vase, I knew it was on the table... If I had known that there were two things on the table to start out with, then when I found one, I might have looked for the other."* (P5)

*6.3.2 Version B: Guided Search.* On average, this task took participants 4.1 minutes (*SD* = 1.9 minutes) to complete. Each object took on average 38.6 seconds (*SD* = 23.3 seconds) to find. Participants generally appreciated the directions, and liked that they updated continuously because it could help correct if they overshot the distance (P10). However, the time interval between directional updates should be customizable, as some participants noted they felt like they were stuck waiting for the app to tell them where to go. P4 noted: *"I don't know if it was me not moving it enough, unless it should be more sensitive. It didn't want to respond right away."* Others suggested adding other multi-modal continuous feedback options to decrease this response time further, for example, P5 suggested a tone that would change as you got closer or further to the object.

### 6.4 Task 4: Furniture Shopping

On average, participants spent 5.8 minutes (*SD* = 1.5 minutes) using this app. Overall participants had positive impressions of the app, noting that it took a while to get used to, but was easy to use after that. Participants generally saw how such an AR app could be useful, but that in its current state, not enough information is provided to base purchasing decisions on.

> *"I liked it, it was cool to be able to place things like that. I guess this would save you from having to measure the furniture, and going to the store and doing that. But sure, it can tell me that the furniture fits in the room, but it doesn't tell me what it looks like once it's placed. The coffee table would fit in front of the couch, but how do I know which way it would fit? How do I know the long side of the coffee table is parallel to the long side of the couch?"* (P9)

Regarding the constituent tasks used in this app, participants had fewer complaints about the camera-based search method being used, presumably because they had a base level of awareness of the locations of objects because they placed them. This is similar to the findings from VizLens [22] suggesting that if users are not familiar with layout, direct guidance is preferred.

However, switching between different tasks was initially confusing to some. For example, our app selected an object when the user stood over it for more than two seconds, as in the object search tasks. As users often walked throughout the space to get a sense of the virtual content, this resulted in unintentional selection. P3 noted at the end of the session *"I was just getting used to it, knowing that I had to step away from objects [to deselect]."* Better consideration of how to integrate our task designs is needed.

Overall, while our prototype is functional and provides some utility, it is missing some of the more intricate information about objects and the environment that would be needed to be fully usable.

### 6.5 Task 5: Educational App

On average, participants spent 8.6 minutes (*SD* = 2.3 minutes) using this app. As previously mentioned, this app is somewhat unique as the layout of the planets was automatically generated, and the users had to find areas of interest. We selected the camera search method for this app with the aim of conveying the exploratory nature of similar educational apps, but it was ill-suited given the unfamiliarity with the layout of the content. Although participants could eventually use the haptic and voice feedback to figure out the arrangement of the planets, it would be less mentally demanding to provide a richer description at the start.

Some participants were unsure of how AR could be useful in this case at all, for example, P10 commented: *"As a blind person, it would be much faster to read about them."* Multiple participants noted that this task would be much more difficult for children, as they would be unable to use prior knowledge about the arrangement of the planets to navigate the app. P2 commented that tactile information should be used as a precursor to virtual information:

> *"For a kid that can't see, bring them something they can touch and feel to give them an idea of the setup. You still need more haptic information. Once they have the setup, they can stop and learn about each planet."* (P2)

## 7 DISCUSSION

Our prototypes allowed users to successfully interact with AR content over a series of tasks. In this section, we discuss overarching patterns that we observed throughout each task regarding users' perceptions of virtual content and additional factors that could affect usability.

### 7.1 Notions of 'Virtual' and 'Real'

Participants interpreted virtual content in various ways, depending on their spatial understanding ability and previous levels of vision. Some participants noted that as blind people already maintain a mental map of an indoor space to some degree, there may be less of a distinction between virtual and physical objects in memory:

> *"Being visually impaired, and being able to see before, now when you have me walk around a room, I had almost exactly in my mind the way the room looks. It's like putting virtual memory in my mind, and to be able to remember [virtual objects], it's almost working in a similar way to actual reality. So it's kinda neat trying to put those two together."* (P3)

Others thought that understanding the virtual environment was much more difficult because the existence of virtual objects created a difference in how the space around them was experienced. P9 commented: *"If I use AR more often, I would come out of my box that I have to touch everything. It's a whole new mindset."* For users who have never had vision, this could be much more difficult, as physical landmarks are the primary way that they understand spatial layout. P7 compared the experience to being in an empty room:

> *"For me, if I'm in a room, and I have a sense of how big the room is, I can get a sense of how furniture is generally laid out; and it's easier to get that concept when the furniture is there. When I was moving in to my last condo, and it was empty, I couldn't conceptualize how everything could go the way I wanted it. For someone who's seeing and has gone blind, they have those spatial concepts, it'll be different for someone like me who has never seen."* (P7)

### 7.2 Richer Contextual Descriptions

Our prototypes focused on providing the user with an awareness of what virtual content was in the environment and the means to interact with it. Given the variations in how participants perceived virtual content, it would likely require less mental effort from the user if future AR applications began to bridge the gap between physical and virtual further. Participants made many comments to this end, motivated by both task performance and safety.

Many participants mentioned wanting an awareness of the physical objects that were relevant to the task at hand (i.e., were in the area that they wanted to place content or could be used as a reference point for an object's location). For example, P9 wanted to know how far they were from the walls when placing an object to center themselves in the room. P1 suggested combining two of our prototypes and providing the user with directions as in guided search to selected physical objects, while they were placing an object with the camera-based placement method. In this manner, the user could be guided to an appropriate area for placement.

Participants also wanted search methods to be aware of physical objects for better navigation, as our prototypes simply instructed users to take the most direct path through the space:

> *"I'm thinking of a situation where there's other things or other people moving around. If the camera's watching that, it might say go forward and put this there. If something runs across in front of you, is it gonna tell you to stop, so you don't crash into them while you're moving a chair."* (P2)

Additionally, such approaches to help users better navigate mixed virtual and physical environments are generally applicable and can also be extended as AR tools and methods for accessible navigation.

## 7.3 Additional Contextual Factors

While we were able to uncover interesting insights on the strengths and weaknesses of each prototype (e.g., unfamiliar layouts made camera search harder) and on user perceptions, some factors were not investigated in our study. For example, different interaction methods may be easier to use depending on the size and complexity of one's physical environment, as well as the user's level of familiarity with the space. Additionally, the location or intended location of a virtual object (on a physical surface, in mid-air, somewhere unreachable, etc.) may also impact the usability of certain interactions. Prior work has shown the scale of virtual objects in AR can have an effect on expected gestural interaction [51], and different interaction techniques have been used to manipulate virtual objects at different distances from the user with success [61]. Similarly, different accessible interaction techniques are needed for these various situations. While virtual objects in our study were mainly placed on the floor and on the desk, we observed some instances of the virtual object's location influencing the usability of the interaction technique. When completing the search task with both prototypes, an object was occasionally placed on the far side of the desk such that the participant had to lean over the desk slightly to select it. While we assumed most locations in the room could be reached fairly easily due to its size, this highlights the need for ergonomics to be considered.

# 8 LIMITATIONS AND FUTURE WORK

In accessibility, we are often in the position of playing "catch up" to make new technologies possible to use. We hope that the work presented here will help set an agenda for creating accessible ways to perform the common interactions necessary in AR. In order to provide fully accessible experiences in this manner, continuing this design work and addressing some of the limitations laid out in the previous section will be important. Our prototypes represent alternatives for only a small number of constituent tasks, and are thus currently best seen as a starting point. For example, our prototypes did not explore animated virtual content or game mechanisms, which are common components present many opportunities for future research. Our prototypes also were created for use with VoiceOver, and thus may not be suitable for people who use other accessibility tools, such as Switch Control, or combinations of tools. However, our hope is that our taxonomy of common AR tasks, as well as the findings and limitations outlined in the previous sections, can serve as a roadmap for our community to explore the large space of AR accessibility comprehensively.

We encourage researchers in this area to take a participatory approach to developing accessible AR. Ideally, this would take the form of engaging with people with disabilities from the start as new AR applications or platforms are developed. The work presented in this paper was a reaction to inaccessible AR applications that have already been released. The goal of our prototypes was to demonstrate that access to visual AR is possible, which we believe our user studies demonstrated, and to show how our taxonomy of interactions could be applied in practice. The particular design decisions were not directly made in consultation with potential users and, thus, should not be seen as necessarily being the best alternative designs for these interactions. A challenge for future work will be to develop AR accessibility that is not only possible but usable and desirable. This will inevitably require early and continuous involvement of the target user group*, e.g., people who are blind*, and will be difficult unless accessibility is considered from the beginning of application and platform design.

Advancing accessible interaction techniques will require not only additional research and design work, but also advances in creating more semantic descriptions of a 3D space from a traditional smartphone camera system. This includes the technical work of creating more robust mapping and tracking systems and generating semantic labels for such maps, and also work in language processing in order to describe scenes [33] and summarize their content [50] to users, and to understand meaningful user requests. Having a richer contextual understanding of users' environments would lead not only to more accessible AR, but better and more natural AR experiences for everyone.

Mixed-reality scenarios in which virtual and physical content deeply interact are especially challenging areas for future work. In some cases, it may make sense to create a virtual version of the physical world that could then be more easily manipulated and consumed alongside the virtual AR content. Virtual reality is easier in some ways to make accessible because the system knows (in theory) about all content in the virtual world. This would be an extension of our "freeze" concept (Section 4). People with motor impairments could move virtually through VR to access location-based AR content, and people with visual impairments could interact with AR content without needing to aim their cameras at a specific point in the real world. Thus, future work could fruitfully explore technologies and interactions for fluidly moving between AR and VR as a useful tool for supporting accessibility.

Finally, as demonstrated in this paper, AR is quickly being adopted across a wide variety of domains, often in a form that considers the visual experience first. With AR and other emerging technologies, we have the opportunity to consider accessibility and multi-modal interactions from the beginning of the design process, rather than as an afterthought [42]. At the same time, history suggests that simply making accessibility features possible for developers to include in their applications will be insufficient, and so future work may usefully explore how to automatically identify AR interactions like those demonstrated in this paper and automatically adapt them to be accessible, thus making AR accessibility ubiquitous. The accessibility research community will undoubtedly be able to draw upon its expertise in accessibility evaluation [9] and fixing [70, 72] in other mediums to apply to accessible AR.

Looking forward, we see great opportunity to go beyond making accessibility possible (necessary and important) and on to using AR technologies for improving accessible experiences. AR is fundamentally about connecting users and our devices to our world, and future work that uses other modalities (e.g., audio, tactile, etc.) as core AR output may open up this rich connection to a wider audience of users.

## 9 CONCLUSION

In this paper, we have presented a taxonomy of tasks that are used in 105 existing mobile AR applications available on the iOS platform. We have created five prototype interactions, and two accessible AR experiences, which served as both design probes and exemplars of accessible alternatives for common AR applications. A study with 10 blind participants demonstrated that our accessible interactions enabled them to use AR applications, and put forth a set of challenges and areas for future research for making AR fully accessible. AR technologies will likely underlie a wide array of new and interesting experiences. Our work provides a path to ensuring that this future is accessible to all.

## ACKNOWLEDGMENTS

We thank our study participants and our reviewers for their time and feedback.